\documentclass[final,1p,times]{elsarticle} 

\usepackage{graphicx}
\usepackage{amssymb} 
\usepackage{amsthm} 
\usepackage{lineno}

\journal{Nuclear Physics A} 

\begin{document}

\begin{frontmatter} 

\title{Influence of initial state fluctuations on the production of thermal photons}
\author{Rupa Chatterjee$^{a}$\footnote{rupa.r.chatterjee@jyu.fi}, Hannu Holopainen$^{a,b,c}$, Thorsten Renk$^{a,b}$, and Kari J. Eskola$^{a,b}$}
\address{$^a$Department of Physics, P.O.Box 35, FI-40014 University of Jyv\"askyl\"a, Finland}
\address{$^b$Helsinki Institute of Physics, P.O.Box 64, FI-00014 University of Helsinki, Finland}
\address{$^c$Frankfurt Institute for Advanced Studies, Ruth-Moufang-Str. 1, D-60438 Frankfurt am Main, Germany}

\begin{abstract}  
Inhomogeneities in the initial QCD matter density distribution increase the production of thermal photons significantly compared to a smooth initial-state-averaged profile in the region $p_T > 1$ GeV/$c$ in an ideal hydrodynamic calculation.  This relative enhancement is more pronounced for peripheral collisions, for smaller size systems as well as for lower beam energies. A suitably normalized ratio of central-to-peripheral yield of thermal photons reduce the uncertainties in the hydrodynamical initial conditions and can be a useful parameter to study the density fluctuations and their size. The fluctuations in the initial density distribution also lead to a larger elliptic flow of thermal photons for $p_T >$ 2.0 GeV/$c$ compared to the flow from a smooth profile.
\end{abstract} 

\end{frontmatter} 

\section{Thermal photons from event-by-event hydrodynamics}
Event-by-event hydrodynamics with fluctuating initial conditions (IC) leads to better agreement with experimental data for particle spectra and elliptic flow of hadrons produced in heavy ion collisions than hydrodynamics with a smooth initial-state-averaged profile~\cite{hannu}. Thermal photons with $p_T > 1$ GeV/$c$ are mostly emitted from the hot and dense early stage of the system expansion. They are thus especially suitable for probing IC fluctuations. For 0--20\% central Au+Au collisions at RHIC, the thermal photon $p_T$ spectrum from fluctuating IC using ideal hydrodynamics is found to be about 10\% flatter than the spectrum from smooth IC in the region $2 \le p_T \le 4$ GeV/$c$.  Consequently, it explains the PHENIX direct photon data better in that $p_T$ range~\cite{chre}. We show that a systematic study of photon observables at different collision centralities and collision energies provides useful information about the density fluctuations and their size in the IC~\cite{chre1}.

The event-by-event hydrodynamic framework developed in~\cite{hannu} is used here (as in~\cite{chre}) to calculate the thermal photon  $p_T$ spectra at different collision centralities at RHIC and LHC energies as well as photon elliptic flow from fluctuating IC. This is a Monte Carlo Glauber based model where the entropy density $s$ is distributed around the wounded nucleons (WN) in the transverse plane using a 2D Gaussian function of the form
\begin{equation}
  s(x,y) = \frac{K}{2 \pi \sigma^2} \sum_{i=1}^{\ N_{\rm WN}} \exp \Big( -\frac{(x-x_i)^2+(y-y_i)^2}{2 \sigma^2} \Big).
 \label{eps}
\end{equation}
 Here, the free parameter $\sigma$ determines the fluctuation size. We use state of the art photon rates (QGP and hadronic rates from~\cite{amy} and~\cite{trg} respectively)  to calculate the photon $p_T$ spectra at RHIC and LHC energies for different collision centralities. The initial formation time $\tau_0$ of the plasma is taken as 0.17 and 0.14 fm/$c$ at RHIC and LHC, respectively, from the EKRT model~\cite{chre1}. 
Results from the fluctuating IC are obtained by taking a final state average of photon $p_T$ spectra from a sufficiently large number of random events. At different centrality bins the smooth IC is constructed by taking an initial-state average of 1000 fluctuating profiles, which essentially removes all the fluctuations from the interior of the fireball.  

\begin{figure}
\centering
\includegraphics[height=4.4cm, clip=true]{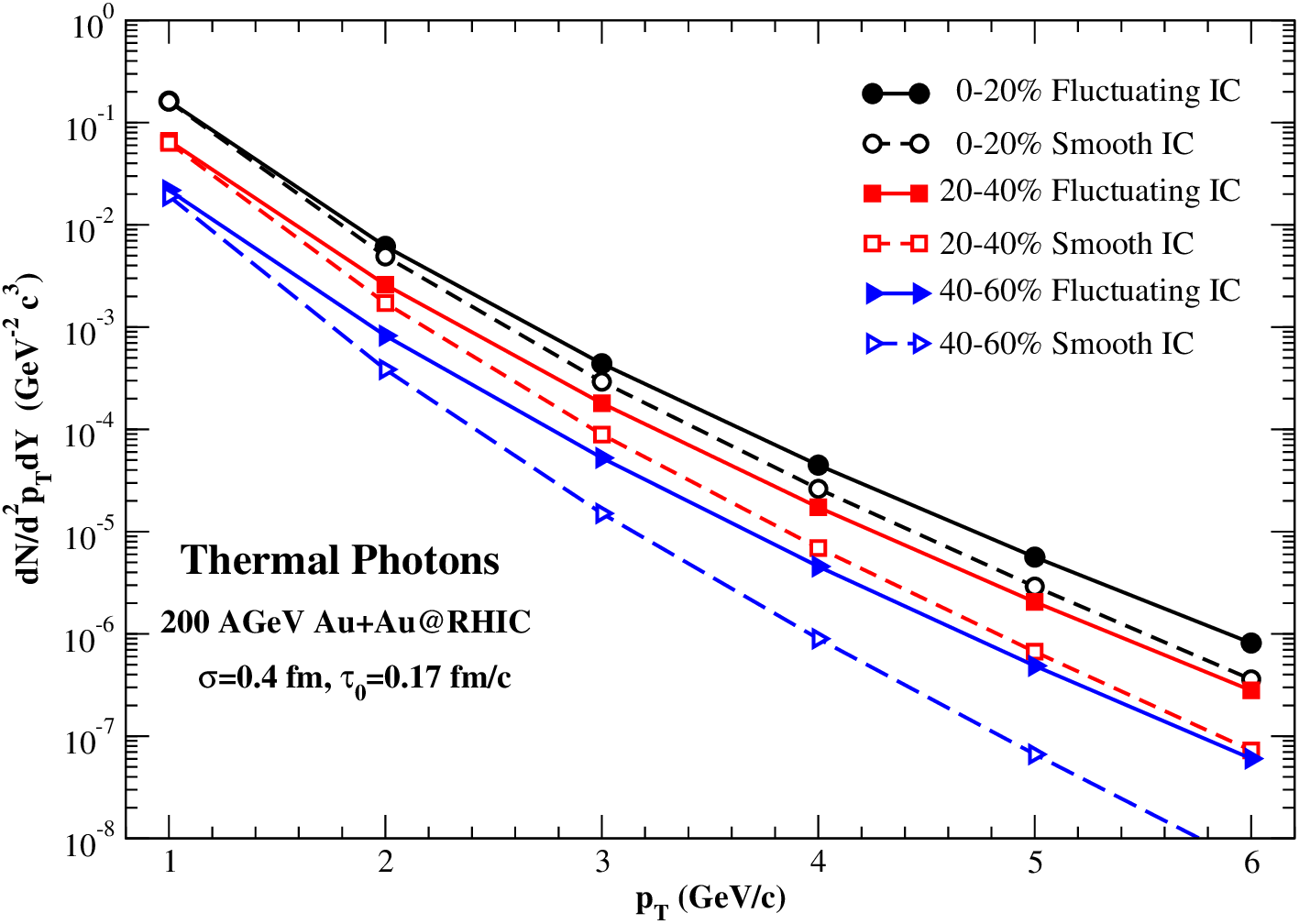}
\includegraphics[height=4.4cm, clip=true]{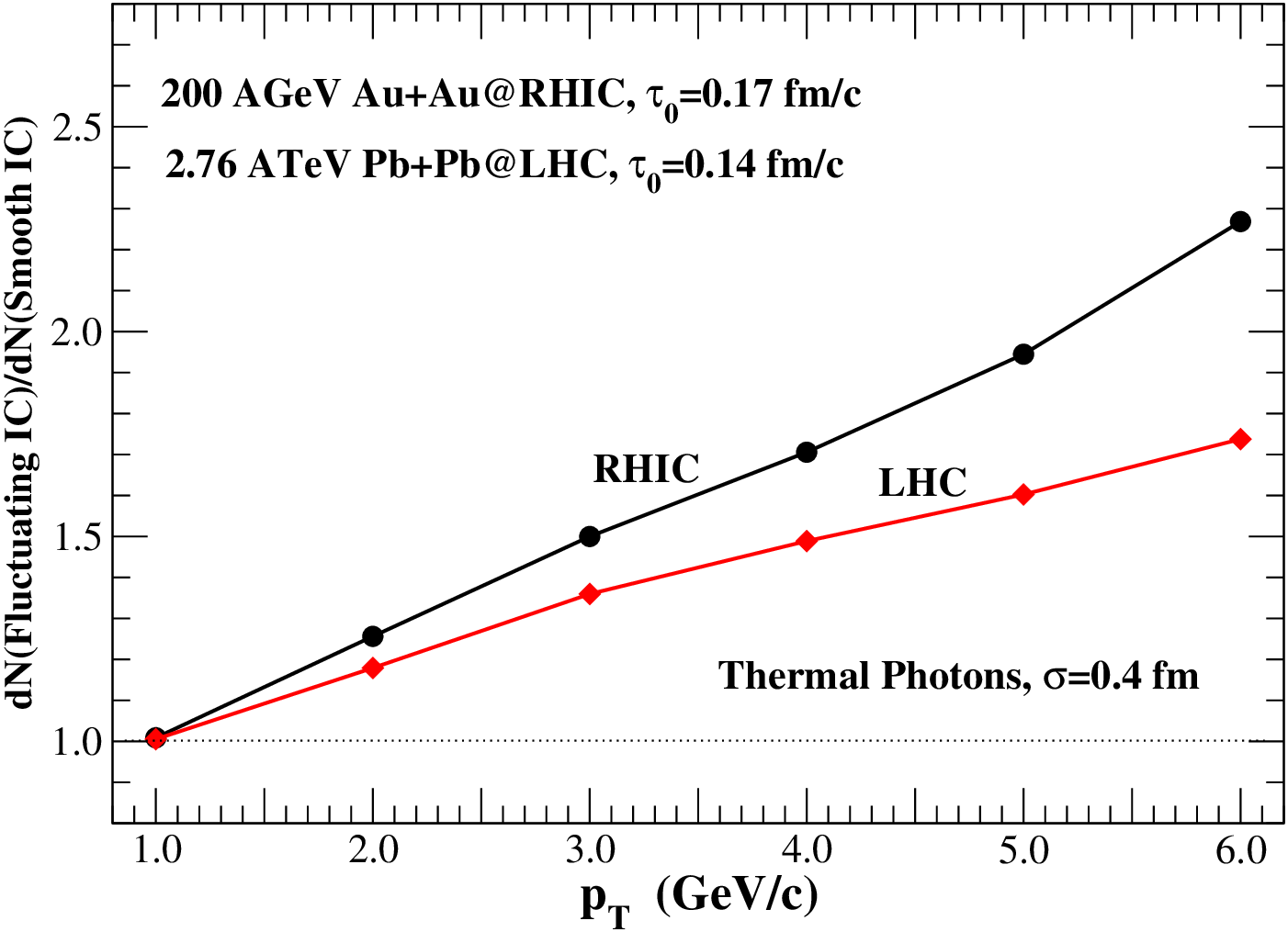}
\caption{[Left] $p_T$ spectra of thermal photons from smooth and fluctuating IC at RHIC for different collision centralities. [Right] Relative enhancement due to fluctuations in the IC at RHIC in comparison with LHC for 0--20\% centrality bin (from~\cite{chre1}).}
\label{fig1}       
\end{figure}

The $p_T$ spectra of thermal photons from Au+Au collisions at RHIC  are shown in the left panel of Figure~\ref{fig1} for three different centrality bins. The results from the smooth and fluctuating IC are obtained using fixed values of $\sigma \, (=0.4 \, \rm {fm})$ and $\tau_0 \, (=0.17 \, \rm fm/c)$ for all the centrality bins. The enhancement in photon production due to IC fluctuations is found to be more pronounced for peripheral collisions than for central collisions, i.e.,  the relative importance of the 'hotspots'  increases for peripheral collisions. The presence of even a few hotspots in the fluctuating IC enhance the photon production significantly as compared to the smooth IC as the photon yield falls rapidly towards peripheral collisions. Similar to RHIC, the effect of the IC fluctuations is also found to be stronger for peripheral collisions than for central collisions at the LHC.  However, the relative enhancement is found to be less at the LHC than at RHIC for the same centrality bin. The ratio of the thermal photon yield from the fluctuating and smooth IC at RHIC and LHC energies for the 0--20\% centrality bin is shown in the right panel of Figure~\ref{fig1}.
\subsection{Initial formation time dependence}
The thermal emission of photons depends strongly on the initial formation time of the plasma. A smaller  $\tau_0$ leads to a larger initial temperature and more high $p_T$ photons from the plasma phase. However, the value of $\tau_0$ is not known precisely and it varies within a range of 0.17 to 0.60 fm/$c$ in different hydrodynamic calculations at RHIC energy. We see that the $p_T$ spectra from smooth as well as from fluctuating IC fall sharply in the range $2 \le p_T \le 4$ GeV/$c$ for all centrality bins when $\tau_0$ is changed from 0.17 fm to 0.60 fm/$c$ (keeping the total entropy of the system fixed, see Figure 4 in ~\cite{chre1}). 
In addition, the value of $\tau_0$  may not remain the same for different centrality bins. The system is more dilute for peripheral collisions and one can expect a larger $\tau_0$ there compared to central collisions. 

\begin{figure}
\centering
\includegraphics[height=4.4cm, clip=true]{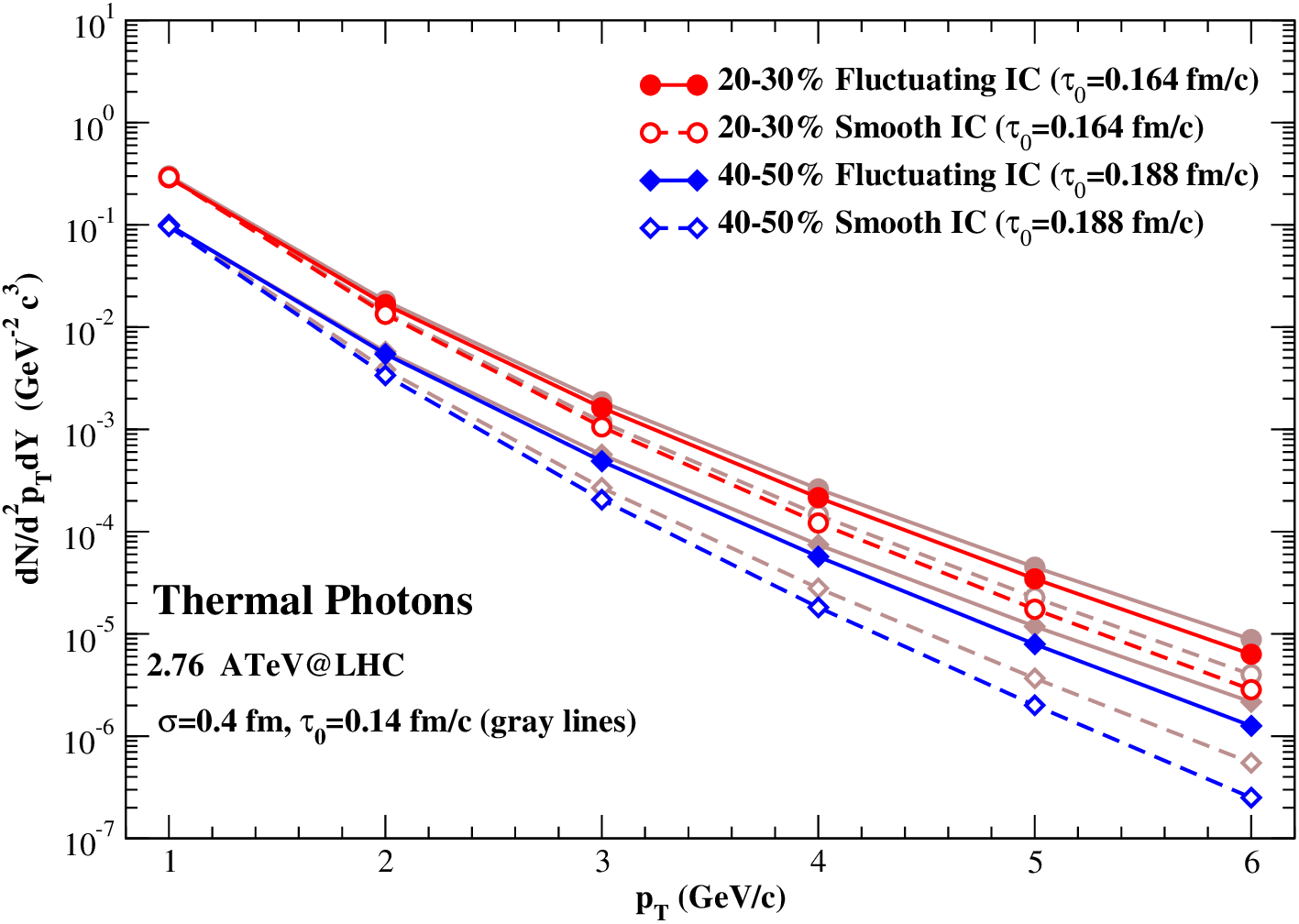}
\includegraphics[height=4.4cm, clip=true]{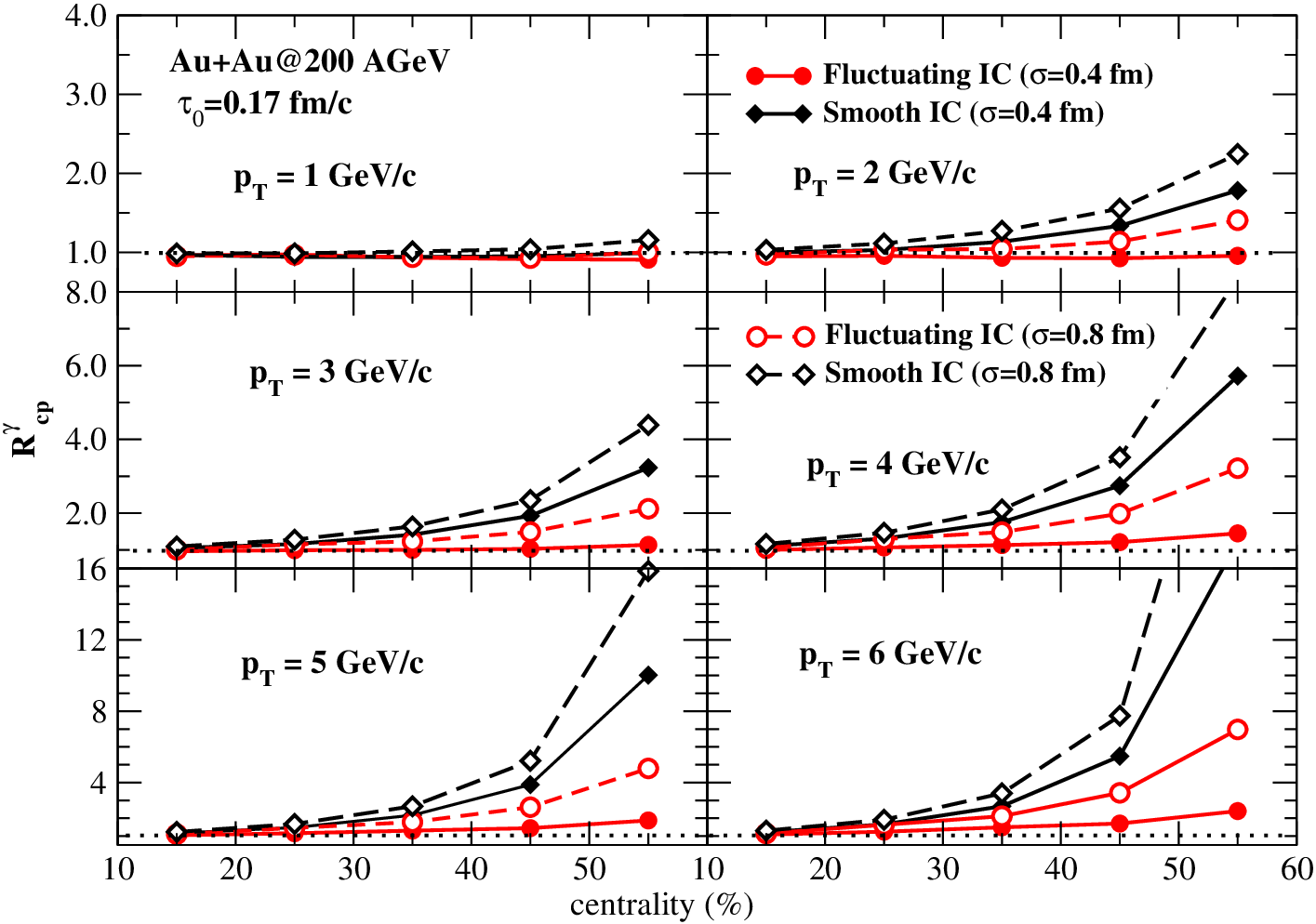}
\caption{[Left] $p_T$ spectra are sensitive to the value of $\tau_0$ which may also vary with collision centrality. [Right] Ratio of central to peripheral yield of thermal photons as a function of collision centrality (from~\cite{chre1}).}
\label{fig2}       
\end{figure}

The left panel of Figure~\ref{fig2} shows $p_T$ spectra of thermal photons at the LHC energy considering centrality dependent $\tau_0$ values (calculated from a non-local version of the EKRT model~\cite{ekrt} with $K_{\rm NLO}^{\rm pQCD}=1$) and fixed $\tau_0$ for all centrality bins (see~\cite{chre1} for details). 
A larger $\tau_0$ for peripheral collisions decreases the production compared to a smaller $\tau_0$, whereas IC fluctuations increase the production. However, due to the uncertainties in the initial formation time it is difficult to conclude anything about the density fluctuations and their size in the IC by looking at the $p_T$ spectra alone.

\subsection{Ratio of central to peripheral yield $R_{\rm cp}^\gamma$}
A suitably normalized ratio of central to peripheral yield can be a useful measure of the fluctuation size parameter by reducing the uncertainties in the initial conditions. 
We define a quantity $R_{\rm cp}^{\gamma}$ as~\cite{chre1}
\begin{equation}
R_{cp}^{\gamma}|_i= \frac {dN/d^2p_TdY|_{0-10 \%}} {dN/d^2p_TdY|_{i-j \%}} \times \frac  {N_{bin}|_{i-j\%}} {N_{bin}|_{0-10 \%}},
\end{equation}
where the value of $i$ is changed from 10 to 70 in steps of 10 and $j$=$i$+10 in the following. For pQCD direct photons (not included here) such a ratio is close to unity. $N_{\rm bin}$ is the number of binary collisions for a particular centrality bin.

The right panel of Figure~\ref{fig2} shows the variation of $R_{\rm cp}^\gamma$ as a function of collision centrality for 200A GeV Au+Au collisions at RHIC and for different values of $p_T$ and $\sigma$. For smaller $p_T$ and for central collisions the value of $R_{\rm cp}$ is close to unity by definition. For $p_T> 2$ GeV/$c$, $R_{\rm cp}^\gamma$ rises rapidly with collision centralities and a significant difference between the  results from the smooth and fluctuating IC can be observed even for larger values of $\sigma$ and $\tau_0$~\cite{chre1}.  
\subsection{Elliptic flow from fluctuating initial conditions}
\begin{figure}
\centering
\includegraphics[height=4.4cm, clip=true]{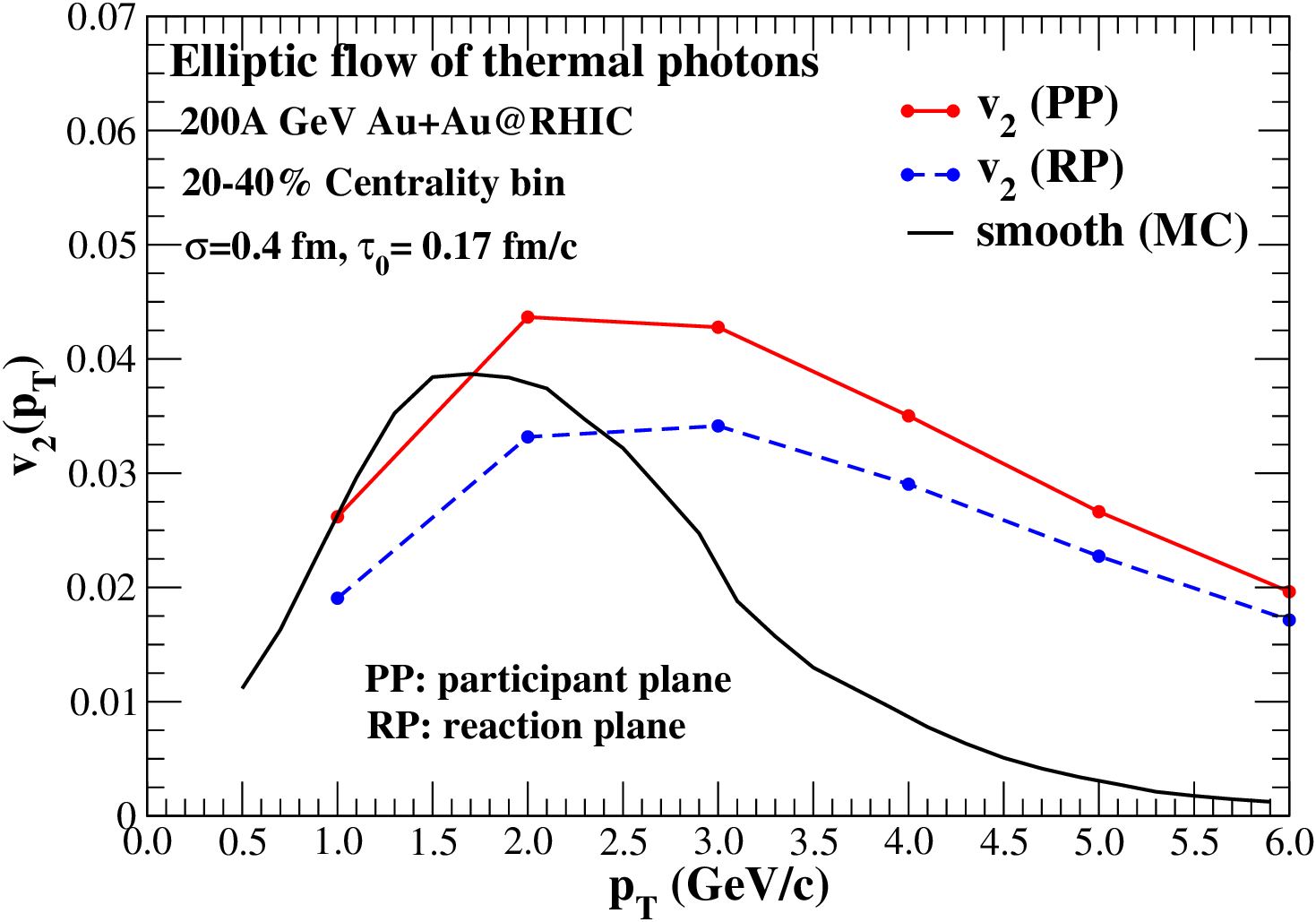}
\includegraphics[height=4.6cm, clip=true]{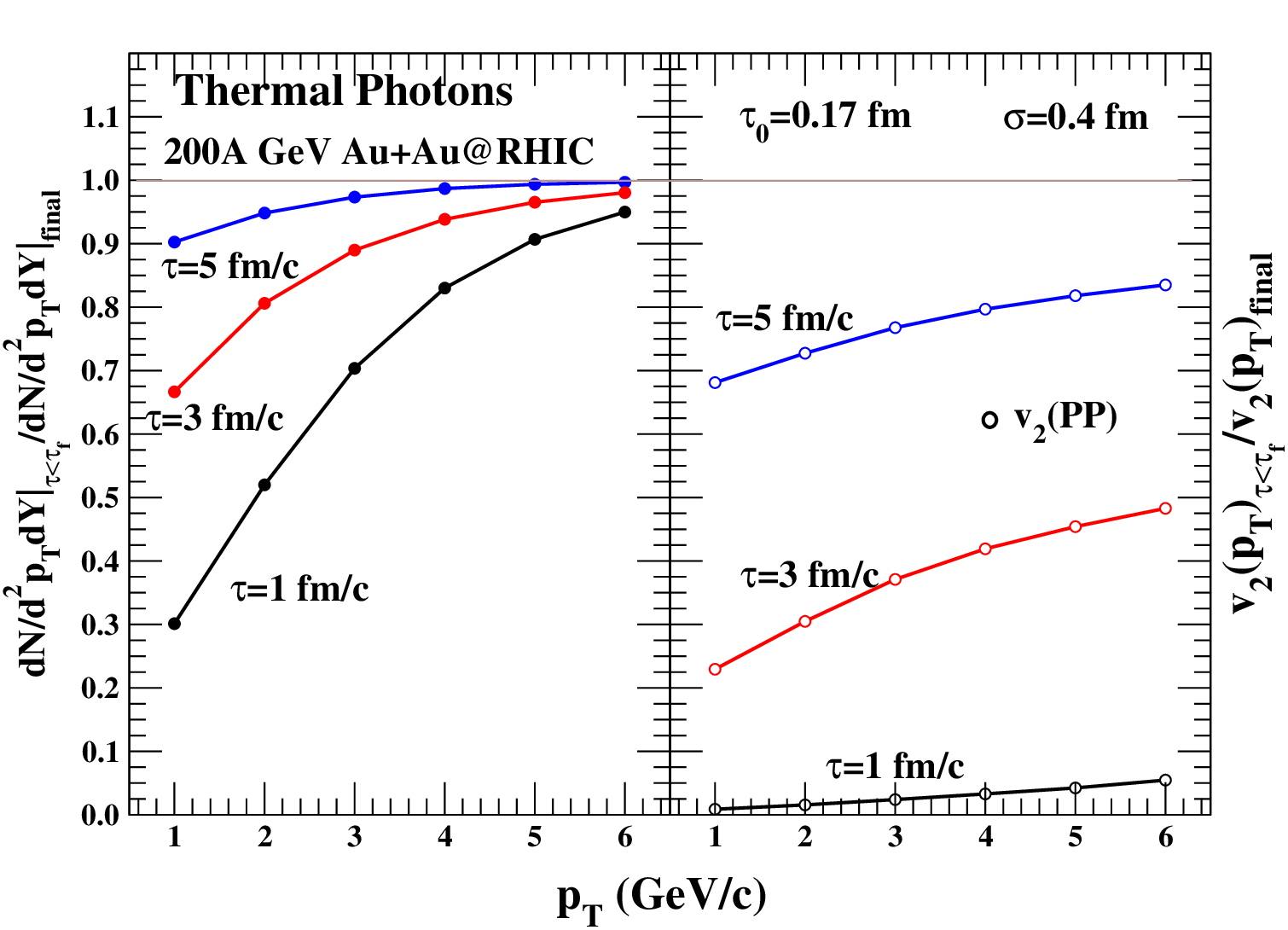}
\caption{[Left] Elliptic flow of thermal photons at RHIC from fluctuating initial conditions. [Right] Time evolution of thermal photon $p_T$ spectra and elliptic flow (normalized by the final $p_T$ spectra and elliptic flow respectively).}
\label{fig3}       
\end{figure}

The elliptic flow of direct photons is believed to be dominated by the flow contribution from thermal radiation up to  
 $p_T \sim 5$ GeV/$c$ as the contributions from all other sources (apart from thermal) are expected to be small in that $p_T$ range.
We calculate the elliptic flow of thermal photons from event-by-event hydrodynamics with respect to the participant plane using the relation $v_2= \langle {\rm cos} \, 2 (\phi-\phi_{\rm PP})\rangle_{\rm event}$, where $\phi_{\rm PP}$ is the participant plane angle~\cite{chre2}. The obtained $v_2(p_T)$ from smooth and fluctuating IC at RHIC for the 20--40\% centrality bin is shown in the left panel of Figure~\ref{fig3}. Fluctuations in the initial density distribution lead to a larger transverse flow velocity and larger elliptic flow especially for $p_T > 2$ GeV/$c$ compared with a smooth profile. However the results from the fluctuating IC are still found to be well below the PHENIX data~\cite{phenix}.
The right panel of Figure~\ref{fig3} shows the time evolution of the thermal photon $p_T$ spectra and elliptic flow (normalized by the final $p_T$ spectra and elliptic flow respectively) of a single event. We see that at $p_T=$ 4 GeV/$c$, more that 80\% of the photons are emitted before first one fm, however less than 10\% of elliptic flow is developed by that time.
\section{Summary and conclusions}
In conclusion, the thermal photon $p_T$ spectra  from smooth and fluctuating IC are compared for different collision centralities at the RHIC and LHC energies. The relative enhancement due to fluctuations in the IC is found to be more pronounced for peripheral collisions and for lower beam energies.
A suitably normalized ratio of central-to-peripheral yield as a function of $p_T$ and collision centrality can be a useful measure of the fluctuation size scale by reducing the uncertainties in the initial condition of a hydrodynamic calculation. Fluctuations in the IC also increase the elliptic flow significantly compared to a smooth profile for $p_T > 2$ GeV/$c$. 
\\

We gratefully acknowledge the financial support by the Academy of Finland. TR and RC are supported by the Academy researcher program (project  130472) and KJE and RC by a research grant (project 133005). HH was supported by the national Graduate School of Particle and Nuclear Physics and currently by the Extreme Matter Institute (EMMI). We acknowledge CSC -- IT Center for Science in Espoo, Finland, for the allocation of computational resources.

\end{document}